\documentclass[prl,showpacs,preprintnumbers,amsmath,amssymb,nofootinbib, superscriptaddress, longbibliography,notitlepage]{revtex4-1}
\usepackage{booktabs}
\usepackage{epsfig}
\usepackage{graphicx}
\usepackage{color}
\usepackage{shuffle}
\usepackage{hyperref}
\usepackage{wrapfig}
\usepackage{cases}
\usepackage[utf8]{inputenc}
\usepackage{mathtools,tabu}
\usepackage{slashed}
\usepackage[title]{appendix}

\def\ZZ{{\mathbb Z}}
\def\cR{{\cal R}}
\def\p{\partial}

\definecolor{dgreen}{rgb}{0,0.75,0.55}
\definecolor{gold}{rgb}{0.85,.66,0}

\usepackage{tikz}
\usetikzlibrary{calc} \usetikzlibrary{patterns,snakes} \usetikzlibrary{decorations.pathreplacing} \usetikzlibrary{decorations.markings} \usetikzlibrary{decorations.pathmorphing} \usetikzlibrary{positioning}

\newcommand\nothing[1]{}

\newcommand\nextstep{Future targets}

\begin{document}

\preprint{UUITP-13/22}
\preprint{NORDITA 2022-015}

\title{Snowmass  White Paper: String Perturbation Theory}


\author{Nathan Berkovits}
\affiliation{ICTP South American Institute for Fundamental Research,
Instituto de F\'isica Te\'orica, UNESP-Universidade Estadual Paulista,
Sao Paulo 01140-070, SP, Brazil}


\author{Eric D'Hoker}
\affiliation{Mani L. Bhaumik Institute for Theoretical Physics, Department of Physics and Astronomy, University of California, Los Angeles, CA 90095, USA}


\author{Michael B. Green}
\affiliation{Department of Applied Mathematics and Theoretical Physics, Cambridge CB3 0WA, UK}
\affiliation{School of Physics and Astronomy, Queen Mary University of London, London, E1 4NS, UK}


\author{Henrik Johansson}
\affiliation{Department of Physics and Astronomy, Uppsala University, 75108 Uppsala, Sweden}
\affiliation{Nordita, Stockholm University and KTH Royal Institute of Technology, SE-10691 Stockholm, Sweden}


\author{Oliver Schlotterer}
\affiliation{Department of Physics and Astronomy, Uppsala University, 75108 Uppsala, Sweden} 

\begin{abstract}
\smallskip
The purpose of this White Paper is to review recent progress towards elucidating and evaluating string amplitudes, relating them  to quantum field theory amplitudes, applying their predictions to string dualities, exploring their connection with gravitational physics, and deepening our understanding of their mathematical structure. We also present a selection of targets for future research.
\end{abstract}

\date{\today}

\maketitle

\section{Introduction}

String theory unifies gravity and Yang-Mills theory in an ultraviolet finite quantum theory that 
is shedding light on a wide range of subjects including particle physics, gauge/gravity duality, black holes, 
gravitational waves, the early universe, and pure mathematics. Several comprehensive 
textbooks on string theory are available \cite{Green:2012oqa, Green:2012pqa, 
Polchinski:1998rr, Polchinski:1998rq, Zwiebach:2004tj, BBS, Blumenhagen:2013fgp, Kiritsis:2019npv}.

\smallskip

String theory may be approached from a number of complementary angles. String perturbation theory is a topological expansion in powers of the string coupling $g_s$ obtained by summing over random fluctuating surfaces. The low-energy expansion in powers of the string scale $\alpha'$ is given by supergravity plus higher order string corrections to supergravity encoded in local effective interactions.  Supergravity is valid for low energy but holds for all values of $g_s$ while string perturbation theory is valid for small $g_s$ but holds for all energies. The gauge/gravity correspondence  formulates string theory in certain hyperbolic space-times via correlators in Yang-Mills theory. At the intersection of these approaches, effective interactions are calculable in string perturbation theory or in the gauge/gravity correspondence, and may be able to predict physical effects observable at low energy. 

\smallskip

Superstring perturbation theory has grown into a rich subject whose study has revealed deep connections with quantum field theory amplitudes, D-branes, gauge/gravity duality, gravitational waves, black holes,  algebraic geometry, and modular forms. At the heart of superstring perturbation theory are superstring amplitudes, which are primarily considered for massless gravitons,  gauge bosons and their respective supersymmetry partners. Following  the discovery of Type~I, Type~II, and Heterotic superstrings in~\cite{Green:1980zg, Green:1981yb, Gross:1984dd}, respectively, 
much of the formalism of superstring perturbation theory dates 
back to the 1980's (see \cite{Friedan:1985ge,DHoker:1988pdl} and references therein for publications prior to 1988). Extensive lecture notes and video lectures on more recent developments may be found in \cite{DHoker:book,Berkovits:2017ldz, ericlect, olilect}.

\section{Perturbative string amplitudes}

The developments of the past 20 years have significantly advanced the explicit computation of a number of string amplitudes, and sharpened our understanding of the formalism. This progress was driven by a confluence of ideas from different formulations of the superstring, including the Ramond-Neveu-Schwarz (RNS)  formalism \cite{DHoker:2001kkt,DHoker:2002hof,DHoker:2005vch} and the pure-spinor formalism \cite{Berkovits:2000fe, Berkovits:2004px, Berkovits:2005bt}. The mathematical structure of the RNS formulation, including the role of supermoduli, was revisited and further clarified in \cite{Witten:2012ga,Witten:2012bh,Witten:2013cia}.
For the pure-spinor formalism, recent work advanced our understanding
of its origin \cite{Berkovits:2015yra} and its interrelation with the
RNS formalism \cite{Berkovits:2016xnb, Berkovits:2021xwh}.

\smallskip

In both formalisms, at one loop and beyond, the chiral splitting procedure \cite{DHoker:1988pdl,DHoker:1989cxq}, which splits the integrands of closed superstring amplitudes  into left and right chiral integrands at fixed loop momenta,  has become a valuable tool in evaluating closed string amplitudes. The structure  of their monodromies constrains the dependence of chiral amplitudes on loop momenta and moduli and thereby significantly simplifies the construction of the amplitudes.  Together with a specification of the D-brane setup, chiral amplitudes also encode the moduli-space integrand for open-string amplitudes.

\smallskip

Tree-level and one-loop amplitudes for four massless supergravity states in the Type~I, Type~II and Heterotic strings were evaluated in the papers that introduced these theories ~\cite{Green:1980zg, Green:1981yb, Gross:1984dd}.  The following selection of more recent benchmark results grew out of progress driven by the development of novel tools for higher loop orders and larger numbers of external gauge and supergravity states.

\smallskip

Tree-level amplitudes are now known in compact and manifestly 
supersymmetric form for any number of massless external legs \cite{Mafra:2011nv}.
The construction was guided by BRST cohomology techniques  and the use of multiparticle 
superfields in the  pure-spinor formalism \cite{Mafra:2014oia}.

\smallskip

Following the early evaluation of the one-loop four-graviton amplitude in \cite{Green:1982sw} and its supersymmetrization in \cite{Berkovits:2004px},  pure-spinor techniques together with chiral splitting
and the multiparticle formalism led to explicit one-loop results up to seven points  (i.e.\ up to seven external states)
\cite{Green:2013bza, Mafra:2016nwr, Mafra:2018qqe}.  In the RNS formalism, the spin-structure sums can be performed at all multiplicities in a controlled function space \cite{Tsuchiya:1988va, Broedel:2014vla}.

\smallskip

The two-loop measure on supermoduli space in the RNS formulation was constructed in \cite{DHoker:2001kkt,DHoker:2001qqx, DHoker:2001foj, DHoker:2001jaf} using the super period matrix, and derived using algebraic geometry   in~\cite{Witten:2013tpa}.
The two-loop amplitude for four massless NS bosons was computed for Type~II and Heterotic strings in the RNS formulation in \cite{DHoker:2005dys,DHoker:2005vch} and was extended to include external fermions in the pure-spinor formulation \cite{Berkovits:2005df, Berkovits:2005ng}. Five-point two-loop amplitudes were constructed from an amalgam of pure-spinor methods and chiral splitting \cite{DHoker:2020prr} in both the Type~II and Heterotic strings, and their parity-even bosonic components were confirmed by a first-principles computation in the RNS formalism \cite{DHoker:2021kks}. Two-loop contributions to the cosmological constant in orbifold compactifications were evaluated in \cite{DHoker:2013sqy, Berkovits:2014rpa}.

\smallskip

At three loops, a measure on supermoduli space in the RNS formulation was proposed in \cite{Cacciatori:2008ay}, further analyzed in \cite{Witten:2015hwa}, and used to discuss  the vanishing of two- and three-point amplitudes in \cite{Grushevsky:2009eqd, Matone:2008td}. The leading low-energy term in the three-loop four-point amplitude was calculated in the non-minimal pure-spinor formalism in \cite{Gomez:2013sla}. Moreover, a recent proposal for the 
four-point amplitude at all energies \cite{Geyer:2021oox} grew out of input from ambi-twistor 
strings \cite{Mason:2013sva, Berkovits:2013xba, Adamo:2013tsa}. Beyond three loops, recent results  include 
proposed non-renormalization theorems \cite{Berkovits:2006vc,Berkovits:2009aw}, and the existence of a super-period matrix on supermoduli space with Ramond punctures \cite{Witten:2015hwa,DHoker:2015gwa}.

\smallskip

The superstring amplitudes described above are given by integrals over the moduli spaces of ordinary compact Riemann surfaces and vertex insertion points on each surface of remarkably simple integrands.  These  integral representations require an analytic continuation in the external momenta. At one-loop order, the analytic continuation was proven  to exist, shown to reproduce all expected physical singularities in terms of a double dispersion relation, and used to compute the decay rate of a large class of massive string states in \cite{DHoker:1993hvl, DHoker:1993vpp, DHoker:1994gnm}.

\smallskip

As will be reviewed in later sections, the remarkable simplicity of the known string amplitudes greatly facilitates deriving gravity and gauge amplitudes from the $\alpha' \rightarrow 0$ limit of the corresponding string amplitudes, verifying  quantitative predictions of dualities, and extracting number-theoretic structures from the low-energy expansion. However, it is not clear whether this remarkable simplicity persists to higher loop order in the RNS formulation due to the increased complexity of supermoduli space \cite{Witten:2012ga,Donagi:2013dua,Sen:2015hia}, and in the pure-spinor formulation due to divergences associated with the composite $b$-ghost \cite{Berkovits:2005bt}.

\smallskip

{\bf \nextstep:} An immediate goal in the RNS formulation is to lift some of the obstacles to higher loop and higher multiplicity amplitudes imposed by the complexity of supermoduli space. Alternatively, one may seek to extract low-energy effective interactions for comparison with S-duality predictions directly from the more involved supermoduli presentation of the amplitudes.  A parallel goal is to systematize and simplify the sums over spin structures, possibly by making use of the novel perspective on the GSO projection developed in  \cite{Kaidi:2019pzj, Kaidi:2019tyf}, and thereby to provide a useful mathematical delimitation for the space of functions to which 
chiral amplitudes belong at arbitrary loop order.

\smallskip

The manifest supersymmetry of the pure-spinor formulation simplifies the evaluation of numerous superstring amplitudes, but higher multiplicities and loop orders are marred by  divergences in the $\lambda,\bar\lambda$-ghost integrals and the difficulties encountered  in evaluating $b$-ghost correlators.  An important goal is to identify the physical origin of  these divergences and to resolve them, possibly building upon ideas in \cite{Aisaka:2009yp, Grassi:2009fe}. 

\smallskip

Another goal is to elucidate further  the relation between the RNS formulation and the pure-spinor formulation. This would lead to a deeper understanding of the superstring, and might allow more general backgrounds for the two formulations. It is not yet known how to describe the RNS formulation in backgrounds with finite non-zero Ramond-Ramond flux, or how to describe the pure-spinor formulation in backgrounds which are not ten-dimensional supergravity solutions. Progress towards understanding this relation between the two formulations has been made in \cite{Berkovits:2016xnb, Berkovits:2021xwh}, 

\smallskip

Perhaps the most daring goal is to obtain higher loop and higher multiplicity amplitudes by indirect methods such as used in bootstrapping conformal field theory or gauge and gravity amplitudes.

\section{Connections with field theory}

Over the past few decades, considerable progress has been made in evaluating and simplifying on-shell scattering amplitudes in gauge and gravity theories (see corresponding White Papers including  \cite{Carrasco}) thereby opening up a wide frontier of results with increasing number of loops and external legs. Supersymmetric theories lend themselves to the greatest degree of simplification and have been studied most extensively in this context. Some of their structures carry over to non-supersymmetric theories. We refer to the textbooks  \cite{Arkani-Hamed:2012zlh, Elvang:2015rqa} and the recent reviews \cite{Bern:2019prr, Travaglini:2022uwo} on modern techniques for quantum field theory (QFT) amplitudes.

\smallskip

Several of these modern techniques were inspired by developments in string theory, such as the relation between supergravity amplitudes and their Yang-Mills counterparts. The intimate relation between open- and closed-string amplitudes, illustrated at tree-level by the Kawai-Lewellen-Tye (KLT) relations \cite{Kawai:1985xq}, and at loop level by chiral splitting  \cite{DHoker:1988pdl,DHoker:1989cxq}, may be viewed as a natural framework for the double-copy structure of supergravity  amplitudes due to Bern, Carrasco and Johansson (BCJ) \cite{Bern:2010ue}.

\smallskip

The BCJ-formulation of the gravitational double copy relies on the so-called color-kinematics duality of its Yang-Mills building blocks \cite{BCJ}. This duality consists of a formal  exchange symmetry of ``color'' degrees of freedom and kinematic variables and can in  many cases be derived from properties of open-string 
amplitudes, as one can see from two complementary perspectives.

\smallskip

First, the manifestly gauge-invariant incarnation of the color-kinematics duality at tree-level in the form of BCJ relations among partial amplitudes \cite{BCJ} stems from monodromy properties of open-string worldsheets 
\cite{BjerrumBohr:2009rd, Stieberger:2009hq}, see 
\cite{Tourkine:2016bak, Hohenegger:2017kqy, Tourkine:2019ukp, Casali:2019ihm, Casali:2020knc} for loop-level generalizations.

\smallskip

Second, the color-kinematics duality can be manifested at the level of cubic-vertex diagrams where kinematic factors obey the same Jacobi identities as color factors. Suitable representations of open-string amplitudes
lead to solutions of kinematic Jacobi identities at tree-level \cite{Mafra:2011kj}, one
loop \cite{Mafra:2014gja, He:2015wgf} and two loops \cite{DHoker:2020prr} through
their respective point-particle limits \cite{Tourkine:2013rda}.
A more general viewpoint on kinematic Jacobi identities is offered by a residue theorem
for moduli spaces of Riemann surfaces with marked points \cite{Mizera:2019blq}
and from nested $b$-ghost action on vertex operators of the pure-spinor formalism 
in Siegel gauge \cite{Ben-Shahar:2021doh}.

\smallskip

Parametrizations of gauge-theory amplitudes subject to kinematic Jacobi identities have shaped the state-of-the-art for multiloop supergravity amplitudes with  different amounts of space-time supersymmetry 
\cite{Bern:2012uf, Bern:2013uka, Bern:2014sna}. Their ultraviolet properties and the associated non-renormalization theorems for gravitational operators can often be anticipated  from low-energy expansions of string amplitudes \cite{Green:2005ba, Green:2010sp, Tourkine:2012ip, Pioline:2018pso}. In particular, studies of the ultraviolet divergences in supergravity  at five-loop order \cite{Bern:2018jmv} were preceded by a corresponding analysis based 
on string theory \cite{Bjornsson:2010wm}.
 
\smallskip

Similar derivations of the gravitational double copy and the color-kinematics duality
of gauge theories at low loop orders have been given within the Cachazo-He-Yuan 
formalism (CHY) \cite{Cachazo:2013gna, Cachazo:2013hca} and ambi-twistor string 
theories \cite{Mason:2013sva, Berkovits:2013xba, Adamo:2013tsa}. These models
are based on worldsheet degrees of freedom similar to those of standard Type~II strings
but their moduli-space integrals are localized via scattering equations, and
dimensionful parameters such as $\alpha'$ are usually absent.

\smallskip

The development of ambi-twistor models and their amplitude prescriptions has strongly benefitted 
from the availability of superstring prototypes. In particular, as exploited in the recent three-loop
proposal for Type-II four-point amplitudes \cite{Geyer:2021oox}, the chiral correlation functions 
of superstrings and ambi-twistor strings can be freely exchanged when expressed in 
terms of logarithmic forms \cite{Gomez:2013wza, Geyer:2018xwu, He:2018pol}.
 
\smallskip
 
Just as string theory elegantly explains far-reaching properties of field theory amplitudes, 
the evaluation of string amplitudes also benefits from field theory structures. For instance, 
the open-superstring tree-level amplitudes themselves admit a double-copy construction 
\cite{Mafra:2011nv} akin to the KLT relations for gravitational amplitudes \cite{Zfunctions}.

\smallskip

Given that the field theory version of the KLT formula usually involves pairs
of physical amplitudes, one may interpret the disk integrals contributing to open
superstrings as scalar amplitudes. Hence, the open superstring is said to be a double copy
of super-Yang-Mills with ``Z-theory'', a putative ultraviolet-soft theory of scalars
with bi-adjoint $\phi^3$ theory and Goldstone bosons in its low-energy
limit \cite{Carrasco:2016ldy, Mafra:2016mcc, Carrasco:2016ygv}.

\smallskip

Similar double-copy constructions apply to bosonic and Heterotic strings \cite{Huang:2016tag, Azevedo:2018dgo}: In their massless tree amplitudes, the super-Yang--Mills building blocks of the open superstring \cite{Mafra:2011nv} are replaced by  the tree-level amplitudes of massive  gauge theories dubbed $(DF)^2+$YM and $(DF)^2+$YM$+\phi^3$ \cite{Azevedo:2018dgo} which have been constructed in the context of conformal-supergravity amplitudes \cite{Johansson:2017srf}.

\smallskip

{\bf \nextstep:} It is an open question whether the double-copy structure of 
supergravity and the color-kinematics duality of gauge theories persist to all orders
in perturbation theory. String theory may offer a suitable framework to investigate
the systematics at higher loops -- either by providing a natural formulation that applies for any
number of loops and legs or by pinpointing fundamental obstructions.

\smallskip

For instance, the five-loop four-point function \cite{Bern:2017ucb} and one-loop six-point function
\cite{Mafra:2014gja, Bridges:2021ebs} of maximal supergravity in general dimensions $D\leq 10$
have so far resisted a double-copy representation in terms of cubic-vertex diagrams. 
While workarounds using the generalized double copy \cite{Bern:2017yxu} and 
linearized Feynman propagators \cite{He:2016mzd, He:2017spx} have been found, it would be 
interesting to check the viability of kinematic Jacobi identities and the traditional
form of the double copy \cite{Bern:2010ue} for these examples from a string-theory perspective.

\smallskip

Another important line of follow-up research concerns double-copy structures
of loop-level string amplitudes. Loop-level KLT relations for closed-string amplitudes
are unknown at the time of writing, and a field theory-type double copy for open-superstring
amplitudes has been pioneered at one loop \cite{Mafra:2017ioj, Mafra:2018qqe}. The
latter calls for a connection with the interpretation of disk integrals as Z-theory amplitudes 
and a systematic understanding in the light of cohomology bases in the chiral-splitting
setting.

\section{String field theory}

Although perturbative superstring theory can be formulated in terms of correlation functions of vertex operators using a two-dimensional worldsheet action, an alternative formulation is as a string field theory in ten space-time dimensions in which the wavefunction of the string field depends on both the zero modes and the excited modes of the worldsheet variables. In contrast to the field theories from the point-particle limit of superstrings discussed in the previous section,
string field theory obtains exact-in-$\alpha'$ expressions for superstring
amplitudes from Feynman-type rules for the string field. Recent lecture notes on string field theory
include \cite{Erler:2019loq, Erler:2019vhl}.

\smallskip

For computing perturbative scattering amplitudes, string field theory is
more complicated than using the worldsheet approach. However, there are several
aspects of superstring theory such as tachyon condensation \cite{Sen:1998sm, Sen:1999nx, Berkovits:2000hf} and mass renormalization \cite{Pius:2013sca, Pius:2014iaa, Sen:2016gqt} which are easier to understand using superstring field theory. Moreover, it is expected that non-perturbative features of superstring theory such as duality symmetries and background independence will be easier to study using superstring field theory. 

\smallskip

Just as there are different formulations of perturbative superstring theory with different worldsheet actions, there are also different formulations of superstring field theory. The best-understood formulation is covariant RNS superstring field theory, and by introducing spurious free fields which decouple from the rest of the action \cite{Sen:2015hha}, it is now possible to describe both the Neveu-Schwarz and Ramond sectors of open and closed superstring field theory \cite{deLacroix:2017lif}. However, space-time supersymmetry is not manifest in this formulation, and ``picture-changing operators" need to be introduced with complicated sewing rules \cite{Sen:2015hia}. Other formulations include light-cone superstring field theory \cite{Green:1984fu}
and WZW-like superstring field theory \cite{Berkovits:1995ab} which have RNS-like versions in which space-time supersymmetry is not manifest, but also have Green-Schwarz-like versions in which
space-time supersymmetry is manifest. There have been several recent works on establishing the relation between these different 
formulations of superstring field theory \cite{Erler:2015uoa, Erler:2017onq, Erler:2020beb}.

\smallskip

{\bf \nextstep:} The relation of open string field theory with closed string field theory needs to be better understood and will probably lead to important new developments. Since open string field theory amplitudes should contain closed string states as poles, open string field theory should also be able to describe closed string fields \cite{Witten:1985cc}. It has been conjectured by many authors \cite{Gopakumar:1998ki,Gaiotto:2003yb,Berkovits:2008qc,Okawa:2020llq} that AdS/CFT duality can be understood as a duality between open string field theory for the CFT side and closed string field theory for the AdS side.

\section{String dualities}

Five superstring theories in 10-dimensional flat Minkowski space-time admit a perturbative expansion in the string coupling, namely Type~I, Type~IIA, Type~IIB, and Heterotic superstrings with gauge groups $E_8 \times E_8$ and Spin(32)$/\ZZ_2$. The low-energy approximation to the sector of massless states in each string theory is governed by the corresponding supergravity. There also exists an 11-dimensional supergravity theory with the maximal number of 32 
supercharges. 

\smallskip

\begin{figure}[htb]
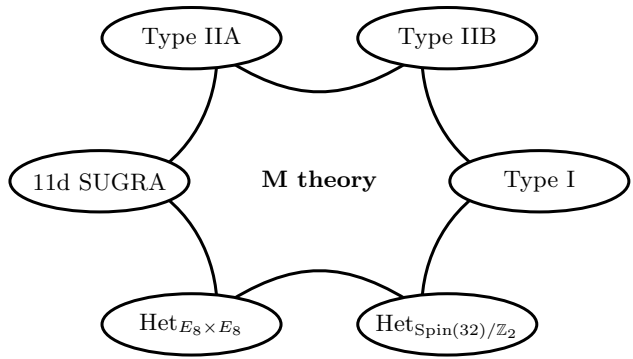

\begin{center}
\tikzpicture[line width = 0.4mm,scale=0.34]
\draw(0,-5)node{{\bf M theory}};
\draw(-4,0) .. controls (-1,-2) and  (1,-2) .. (4,0) ;
\draw(-4,0) .. controls (-4,-2) and  (-5,-4) .. (-7,-5) ;
\draw(-4,-10) .. controls (-4,-8) and  (-5,-6) .. (-7,-5) ;
\draw(-4,-10) .. controls (-1,-8) and  (1,-8) .. (4,-10) ;
\draw(4,-10) .. controls (4,-8) and  (5,-6) .. (7,-5) ;
\draw(4,0) .. controls (4,-2) and  (5,-4) .. (7,-5) ;
\draw[fill=white](-5,0.6) ellipse(3.5cm and 1.2cm);
\draw(-5,0.6)node{Type IIA};
\draw[fill=white](5,0.6) ellipse(3.5cm and 1.2cm);
\draw(5,0.6)node{Type IIB};
\draw[fill=white](8.6,-5) ellipse(3.5cm and 1.2cm);
\draw(8.6,-5)node{Type I};
\draw[fill=white](5,-10.6) ellipse(3.5cm and 1.2cm);
\draw(5,-10.7)node{Het$_{{\rm Spin}(32)/{\ZZ_2}}$};
\draw[fill=white](-5,-10.6) ellipse(3.5cm and 1.2cm);
\draw(-5,-10.6)node{Het$_{E_8 \times E_8}$};
\draw[fill=white](-8.6,-5) ellipse(3.5cm and 1.2cm);
\draw(-8.6,-5)node{11d SUGRA};
\endtikzpicture
\caption{\textit{The web of string dualities, see for instance \cite{Polchinski:1998rq, BBS}.}
\label{figdualities}}
\end{center}
\end{figure}

String dualities are discrete transformations that relate these different theories \cite{Hull:1994ys, Witten:1995ex}, often after suitable compactification. The conjectured existence of these dualities has led to posit that the five perturbative string theories are but different manifestations of a single unifying  ``M-theory". A schematic illustration is given in  Figure~\ref{figdualities}.  We refer to  \cite{Freedman:2012zz} for a modern textbook on supergravity; to \cite{Duff:1999ys} for  a collection of essays on M-theory; to \cite{Obers:1998fb} for an early overview on dualities, and to \cite{Polchinski:1998rq,BBS} for  textbooks with extensive discussions of string dualities.

The large web of dualities between superstring theories  includes  perturbative  as well as non-perturbative dualities. T-duality relates string theories compactified on different tori to one another, such as Type~IIA to Type~IIB,  and Heterotic $E_8 \times E_8$ to ${\rm Spin}(32)/\ZZ_2$, and  gives rise to arithmetic duality groups and associated automorphic forms (see \cite{Obers:1999um,Green:2010kv,Fleig:2015vky} and references therein). Mirror symmetry similarly relates string theories compactified on different Calabi-Yau manifolds. For the vast literature on these subjects we refer to \cite{Hubsch:1992nu,Blumenhagen:2006ci, Yau1, Yau2, Hori:2003ic}, and references therein. Compactification of Type~II on a K3 Calabi-Yau manifold is dual to Heterotic theories on the 4-torus \cite{Hull:1994ys, Vafa:1995fj, Antoniadis:1997eg, Kiritsis:2000zi}, while Type~I theory is dual to the Heterotic ${\rm Spin}(32)/\ZZ_2$  theory \cite{Witten:1995ex, Polchinski:1995df,Tseytlin:1995fy, Tseytlin:1995bi, Bachas:1997mc, Stieberger:2002fh, Stieberger:2002wk}.

\smallskip

Non-perturbative  S-dualities map the weak-coupling regime of one theory to the strong-coupling regime of another or the same theory. Proving the existence of an S-duality is notoriously difficult if not, at present, out of reach. Nonetheless, string perturbation theory can be used to provide indirect, yet powerful, tests of some of the implications of S-dualities that usually relate different loop orders. We shall now illustrate such tests and offer future targets.

\smallskip

One of the best studied non-perturbative dualities  is the $SL(2,\ZZ)$ self-duality of Type~IIB  string theory. 
The low-energy expansion of  string theory, in powers of $\alpha'$,  consists of supergravity plus local effective interactions produced by integrating out the tower of massive string modes. The coefficients of these effective interactions depend on the complexified axion-dilaton field $\Omega$ whose imaginary
part sets the string coupling $g_s$. Self-duality requires the coefficients to be modular forms in $\Omega$ under $SL(2,\ZZ)$, while supersymmetry imposes differential equations on these modular forms.  Remarkably, $SL(2,\ZZ)$ duality and supersymmetry combine to  imply relations between different  perturbative contributions to various effective interactions, and it is these predictions that can be tested against predictions from perturbative string theory.

\smallskip

The lowest order in $\alpha'$ corrections to Type~II supergravity contain the $\cR^4$ effective interaction evaluated at tree-level in \cite{Gross:1986iv,Grisaru:1986px}, where $\cR^4$ symbolically represents a quartic scalar combination of the Riemann tensor whose tensor structure is fixed by supersymmetry. Combining $SL(2,\ZZ)$ duality, D-instanton effects, and supersymmetry, the coefficient of $\cR^4$ was found to be the non-holomorphic Eisenstein series $E_{{3 \over 2}}(\Omega) $  \cite{Green:1997tv, Green:1997as,Green:1998by}. The modular form  $E_{{3 \over 2}}(\Omega)$ produces perturbative contributions at tree and one-loop levels only, and predicts the vanishing of all higher-order perturbative contributions. This prediction was confirmed at two loop order in  \cite{DHoker:2005vch} and  at three loop order in \cite{Gomez:2013sla}. These non-renormalization effects may be attributed to the fact that $\cR^4$ is BPS-protected.

\smallskip

Higher-order corrections in the $\alpha'$ expansion to Type~II supergravity contain $D^{2k} \cR^\ell$ effective interactions for $k \geq 0$ and $\ell \geq 4$, where $D$ is the  covariant derivative. The tree-level contributions to these effective interactions may be obtained by expanding the tree-level closed-string amplitudes in powers of $\alpha'$. As will be detailed in the next section, these expansions can be imported at all multiplicities from  the algorithmic expansion of open-string amplitudes \cite{Broedel:2013aza, Mafra:2016mcc} in terms of multiple zeta values (MZVs) \cite{Terasoma, Brown:2009qja, Schlotterer:2012ny}. Going beyond perturbation theory, D-instanton effects have been further analyzed in \cite{Balthazar:2019rnh, Sen:2020cef, Sen:2020ruy, Sen:2021qdk, Sen:2021tpp, Sen:2021jbr, Alexandrov:2021shf, Alexandrov:2021dyl}.

\smallskip

Combining the tree-level contributions to BPS protected operators at higher order in $\alpha'$ with $SL(2,\ZZ)$ duality, supersymmetry, and D-instanton effects, the coefficient of  $D^4 \cR^4$ was found to be  $E_{{5 \over 2}}(\Omega)$ \cite{Green:1999pu}, while the coefficient of $D^6 \cR^4$ satisfies a modular differential equation \cite{Green:2005ba,Green:2014yxa, Bossard:2020xod}. The structure of these modular forms again predicts non-trivial relations between various perturbative corrections, or their absence. One-loop contributions to $D^{2k} \cR^4$ and $D^{2k-2} \cR^5$ for $k\leq 6$ were extracted from the  four- and five-point  amplitudes in \cite{Green:1999pv, DHoker:2015gmr, DHoker:2019blr} and \cite{Richards:2008jg, Green:2013bza, Basu:2016mmk}, respectively, and successfully matched against the predictions of $SL(2,\ZZ)$ duality. Two-loop perturbative contributions were evaluated and matched as follows:  $D^4\cR^4$ in \cite{DHoker:2005jhf,Gomez:2010ad};  $D^6 \cR^4$ in \cite{DHoker:2013fcx,DHoker:2014oxd};  $D^2 \cR^5$  in \cite{Gomez:2015uha};  and $D^4 \cR^5$  in \cite{DHoker:2020tcq}. The three-loop contribution to $D^6 \cR^4$ was matched in \cite{Green:2005ba,Gomez:2013sla}.

\smallskip

{\bf \nextstep:} An immediate target is to analyze the implications of $SL(2,\ZZ)$ duality in Type IIB on the axion-dilaton field dependence of the  coefficients of low-energy effective interactions that are less or not BPS-protected, such as $D^{2k} \cR^4$ for $k \geq 4$ and terms with higher powers of the curvature. Tree-level results for $D^{2k} {\cal R}^\ell$ with $\ell \geq 5$  already predict the appearance of (single-valued) MZVs of arbitrary depth.  In the low-energy asymptotics of the corresponding string amplitudes at loop level, these operators are invariably accompanied by non-local contributions, due to the exchange of massless states in loops, which complicates their precise definition. Although supersymmetry non-renormalization theorems are expected to apply with less predictive power, early studies suggest the presence of non-trivial mathematical structure \cite{Green:2006gt,Green:2008uj,Green:2008bf,Berkovits:2009aw, DHoker:2017pvk, DHoker:2018mys,DHoker:2019blr}. Progress on such non-renormalization theorems would also help address questions regarding  the ultraviolet properties of 
supergravity  in four space-time dimensions.

\smallskip

A closely related issue is as follows. In spite of detailed control over the tensor structure of open- and closed-string tree amplitudes and their $\alpha'$-expansions \cite{Mafra:2011nv, Schlotterer:2012ny, Broedel:2013aza}, the explicit form of the associated  interactions, such as $D^{2k}R^\ell$,  is unknown beyond the order of $(\alpha')^4$. An important goal is  to develop a method by which the simplicity of string tree-level  amplitudes can be used to construct the detailed tensor structure of the corresponding low-energy effective interactions (see for instance \cite{Koerber:2002zb, Barreiro:2012aw} 
for progress in taming the complexity of traditional methods).

\section{Gauge/gravity duality}

The AdS/CFT correspondence conjectures the quantum equivalence between Type~IIB superstring theory on the $AdS_5 \times S^5$ space-time and ${\cal N}=4$ super-Yang-Mills theory (SYM)  with gauge group $SU(N)$ and coupling $g$ on four-dimensional flat space-time \cite{Maldacena:1997re}. The conjecture provides an explicit relation between on-shell scattering amplitudes in Type~IIB superstring theory and conformal correlators of gauge-invariant operators in ${\cal N}=4$ SYM \cite{Witten:1998qj,Gubser:1998bc}. The large $N$ limit at fixed `t Hooft coupling $\lambda = g^2N$ corresponds to classical superstring theory  on $AdS_5 \times S^5$ while the further limit of large $\lambda$ corresponds to the supergravity limit as the radii of both $AdS_5 $ and $ S^5$ tend to infinity. 
It is especially the correspondence involving the supergravity limit  (including field theories with  less or no supersymmetry and without conformal or Poincar\'e symmetry) that has spawned a record volume of research over the past 25 years (see \cite{Aharony:1999ti, DHoker:2002nbb, Beisert:2010jr} for reviews and \cite{Ammon:2015wua, Nastase:2015wjb} for textbooks on the subject). In this section, we focus on progress towards the quantization of Type~IIB strings on AdS backgrounds, and the interplay between conformal correlators in strongly coupled ${\cal N}=4$ SYM and $SL(2,\ZZ)$ duality. A separate White Paper is dedicated to progress  \cite{Gopakumar:2022kof} on establishing the map between conformal correlators and string perturbation theory.

\smallskip

One of the great challenges of perturbative string theory is to quantize strings in space-times other than flat 
ten-dimensional Minkowski space-time or  toroidal compactifications thereof. This problem is unsolved, even  at string tree-level on space-times with large isometry groups that preserve maximal supersymmetry such as $AdS_5 \times S^5$. A~key obstacle in the RNS formulation is the difficulty of incorporating Ramond flux backgrounds \cite{Metsaev:1998it}. Progress has been made for  backgrounds with pure NS-NS background fields, such as string theory on $AdS_3 \times S^3$ with NS-NS flux through the sphere \cite{Maldacena:2001km,Maldacena:2000kv,Maldacena:2000hw,Gaberdiel:2011vf}. 
To describe the superstring in an $AdS_5\times S^5$ background with finite
radius, one needs to use either the Green-Schwarz \cite{Metsaev:1998it} or
pure-spinor formalism \cite{Mazzucato:2011jt}. Although more computations
have been performed using the Green-Schwarz formalism, the pure-spinor
formalism has the advantage of allowing quantization in a manner that
preserves the full $PSU(2,2|4)$ symmetry \cite{Berkovits:2004xu}.

\smallskip

To compute $AdS_5\times S^5$ scattering amplitudes in which the external states have large $R$-charge, it is convenient to gauge-fix to a light-cone version of the Green-Schwarz superstring in which $PSU(1,1|2)\times PSU(1,1|2)$ symmetry is manifest. In the limit of infinite $R$-charge, the worldsheet action is quadratic and describes a plane-wave background \cite{Metsaev:2001bj}, and numerous tests of the gauge-gravity duality have been performed by expanding around this background (see references in the reviews of \cite{Tseytlin:2010jv, McLoughlin:2010jw, 
Magro:2010jx,Schafer-Nameki:2010qho}).

\smallskip

When the $AdS_5$ radius is small, the superstring amplitudes should coincide with perturbative SYM correlators and one can try to explicitly prove the $AdS_5/CFT_4$ correspondence. Using the hybrid formalism for the superstring in $AdS_3\times S^3$ \cite{Berkovits:1999im} at small radius, the $AdS_3/CFT_2$ correspondence was proven in \cite{Eberhardt:2019ywk} and it was shown that similar methods might be possible for the $AdS_5/CFT_4$ case \cite{Gaberdiel:2021jrv}. Although less is known about the $AdS_5\times S^5$ superstring at small radius, it has been conjectured that the $AdS_5\times S^5$ superstring at zero radius is described by a topological string theory \cite{Berkovits:2008ga} and preliminary evidence for this conjecture was presented in \cite{Berkovits:2019ulm}.

\smallskip

The study of strongly coupled conformal field theories in various dimensions is of central importance for many areas of theoretical physics.   The prime example is ${\cal N}=4$ SYM, whose exact conformal invariance is guaranteed by supersymmetry, whose $SL(2,\ZZ)$ duality is well-established, and whose correlators  are  related by holography to  on-shell amplitudes  in Type~IIB string theory on $AdS_5 \times S^5$.  Here, we shall focus on recent results, obtained by supersymmetric localization techniques, that provide the realization of $SL(2,\ZZ)$ duality on certain {\it integrated correlators}  at arbitrary coupling. 
   
\smallskip

The starting point for the study of integrated correlators is the fact that ${\cal N}=4$ SYM theory can be expressed as a limit of the ${\cal N}=2^*$ theory in which the hyper-multiplet mass $m$ vanishes.  Generalizing results of \cite{Nekrasov:2002qd, Nekrasov:2003rj}, supersymmetric localization techniques were used in \cite{Pestun:2007rz}  to determine the  partition function $Z_{G}(\tau,\bar \tau; m)$  for ${\cal N}=2^*$ SYM on $S^4$  with any classical gauge group $G$ and complex gauge coupling $\tau$. In \cite{Binder:2019jwn} the quantity   $({\rm Im}  \tau)^2 \p_\tau \p_{\bar \tau} \partial_{m}^2 Z_G(\tau, \bar \tau; m)|_{m=0}$ was shown to give the correlator of four superconformal primaries in the stress-tensor multiplet of ${\cal N}=4$ SYM for gauge group $G=SU(N)$,  integrated over their positions with a specific measure that  preserves supersymmetry.  Although detailed information about the space-time dependence is sacrificed by this integration the high degree of supersymmetry of such integrated correlators enables their exact evaluation. The large $N$ limit for fixed 't Hooft coupling $\lambda=g^2 N$ is holographically dual to the perturbative expansion of the Type~IIB superstring  amplitude in $AdS_5\times S^5$ \cite{Binder:2019jwn,Chester:2019pvm}. The analysis is facilitated by using Mellin transform techniques  \cite{Mack:2009mi,Penedones:2010ue}  and was generalized to an arbitrary classical gauge group $G$ in  \cite{Alday:2021vfb}.

 \smallskip
 
While Yang-Mills instantons are suppressed in the large $N$ expansion at fixed `t Hooft coupling $\lambda = g^2 N$, they are unsuppressed in the  large $N$ expansion for fixed $g^2$ considered in \cite{Chester:2019jas}. The successive powers of $1/N$ now correspond to successive  terms in the low-energy expansion of the Type~IIB amplitude. The coefficient of each power is a sum of non-holomorphic $SL(2,\ZZ)$ Eisenstein series and  Montonen--Olive duality \cite{Montonen:1977sn, Witten:1978mh,  Osborn:1979tq} is manifest.    This is the gauge theory image of $SL(2,\ZZ)$ S-duality of Type~IIB superstring theory.    Combined with information gathered from other integrated correlators \cite{Chester:2020dja,Chester:2020vyz}, these results  reproduce the known coefficients of the ${\cal R}^4$ and $D^4 {\cal R}^4$ effective interactions, but also contain sub-leading corrections that are suppressed by powers of the $AdS_5$ length scale.  

\smallskip

Extracting more detailed information from the partition function  $Z_{SU(N)}(\tau, \bar\tau;m)$ is complicated by the complexity of the multi-instanton sectors. However, in  \cite{Dorigoni:2021guq}  an elegant reformulation of the  integrated correlator was conjectured in terms  of a two-dimensional lattice sum, that is well-defined for all values of $N$ and $\tau$,  may be expressed formally as an infinite sum of non-holomorphic Eisenstein series,  has a simple spectral representation \cite{Collier:2022emf},   and generalizes to any classical gauge group  \cite{Dorigoni:2022zcr}.  For groups other than $SU(N)$  Montonen-Olive duality  extends to include Goddard-Olive-Nuyts duality \cite{Goddard:1976qe}, which relates a theory with gauge group $G$ to a theory with the Langlands dual gauge group, ${}^LG$.  Furthermore, for  any classical gauge group $G$ the integrated correlators  satisfy  Laplace difference equations that relate them  to the integrated correlators for $SU(2)$.    For classical groups other than $SU(N)$ the background geometry of the  holographically dual string theory  is an orientifold associated with $AdS_5\times S^5/\ZZ_2$   \cite{Witten:1998xy}.

\section{Gravitational waves and black holes}

The detection of gravitational waves from binary systems composed of black holes and/or neutron stars by the LIGO/Virgo collaboration~\cite{LIGOScientific:2016aoc,LIGOScientific:2017vwq} has initiated a new era for studying the cosmos. Gravitational phenomena are again at the core of fundamental physics, both in experiments and theory, and we are heading towards a bright high-precision frontier. Highly precise analytic calculations in general relativity are indispensable to build waveform models for the inspiral phase of binary mergers, and thus infer unique astrophysical and fundamental physics information hidden in the gravitational-wave signals.   Remarkably, QFT and string theory provide advanced mathematical tools associated to scattering amplitudes that display stunning simplicity when applied to the non-linear problem of classical gravity~\cite{Luna:2016due,Bjerrum-Bohr:2018xdl, Shen:2018ebu, Cheung:2018wkq, Kosower:2018adc, Bern:2019nnu, Blumlein:2019zku, Maybee:2019jus, Johansson:2019dnu, Damgaard:2019lfh, Bern:2019crd, Kalin:2019rwq,Kalin:2019inp, Levi:2020kvb, Chung:2020rrz, Cheung:2020gyp, Cheung:2020sdj, Cristofoli:2020uzm, Bern:2020buy, Kalin:2020mvi, Kalin:2020fhe, Levi:2020lfn, DiVecchia:2020ymx, Bern:2020uwk,Mogull:2020sak, Cristofoli:2020uzm,  AccettulliHuber:2020dal,  Bern:2021dqo, DiVecchia:2021bdo, Brandhuber:2021kpo, Dlapa:2021npj, Damgaard:2021ipf,  Cristofoli:2021vyo, Brandhuber:2021eyq,  Chen:2021qkk,   Kim:2021rfj, Cho:2021arx, Bern:2021yeh, Jakobsen:2022fcj, Veneziano:2022zwh, Edison:2022cdu,Manohar:2022dea}. A separate White Paper~\cite{GW_ampl_WP} is dedicated to the emerging subfield of applying modern amplitudes methods to the problem of gravitational-wave physics, see further references therein. 

\smallskip

With higher sensitivity LIGO, Virgo and KAGRA observing runs in the next few years, and with upcoming detectors (Einstein Telescope, Cosmic Explorer, LISA) it is crucial to further improve the accuracy of theoretical predictions.  Even with state-of-the-art analytical effective field theory (EFT) methods the needed post-Newtonian (PN) or post-Minkowskian (PM) calculations are highly challenging. The complexity can be traced back to the sheer number of terms appearing in the gravitational Feynman diagrams, as well as to the problem of analytic integration. For the integrals, the challenge is ameliorated by the increased understanding of integral bases, integration by parts identities, function spaces graded by transcendental weight, and differential equations --  tools originally developed for QFT and string scattering amplitudes.  The complexity of the integrands can be managed by exploiting the double-copy relationship between gravity and gauge theory -- originally in the form of the KLT relations between open- and closed-string amplitudes. In combination with unitarity-based methods~\cite{Bern:1994zx, Bern:1994cg,Britto:2004nc}, this has launched a revolutionary path to access precision gravity via vastly simpler gauge theory calculations, recently reaching new heights at 3PM and 4PM orders~\cite{Bern:2019nnu,Bern:2021dqo} for Schwarzschild black holes.

\smallskip

The PM calculations are necessary for describing scattering events of pairs of unbound black holes to all orders in the velocity,
and they provide vital cross-checks and simplifications to the PN calculations needed for bound systems. Scattering of black holes benefits from the eikonal approach~\cite{Ciafaloni:2018uwe,DiVecchia:2021bdo}, which in string theory was explored at high energy in the Amati-Ciafaloni-Veneziano (ACV) formalism~\cite{Amati:1987wq,Amati:1987uf}. The latter approach recently led to the resolution of an open question with the conservative dynamics at 3PM~\cite{DiVecchia:2020ymx}. Further, tangentially related QFT approaches based on Weinberg's soft theorem~\cite{Weinberg:1964ew}, gravitational memory~\cite{Zeldovich:1974gvh} and celestial holography (see e.g.~\cite{Strominger:2017zoo,Raclariu:2021zjz,Pasterski:2021rjz, Pasterski:2021raf} and references therein), have lead to novel perspectives on gravitational wave emission~\cite{Laddha:2018rle,Addazi:2019mjh,DiVecchia:2021ndb}.

\smallskip

Properly incorporating spin effects in general relativity is paramount for mergers involving astrophysical Kerr black
holes, with accurate parameter estimation. 
In an EFT approach~\cite{Goldberger:2004jt,Porto:2005ac}, the spin multipole moments are
included to match results from black hole perturbation theory and self-force calculations, but vast theoretical challenges remain. For a classical Kerr black hole, there is currently no satisfactory EFT action valid for higher-order calculations in both the gravitational coupling and spin. The worldline EFT approach to Kerr encounters challenges starting with unknown Wilson coefficients of the Riemann-squared operators at order $S^4$~\cite{Siemonsen:2019dsu}, where $S$ is the classical spin vector. Likewise the Compton scattering amplitude for a putative Kerr black hole develops spurious poles beyond quantum spin $s>2$~\cite{Arkani-Hamed:2017jhn,Guevara:2018wpp,Arkani-Hamed:2019ymq}. 

\smallskip

{\bf \nextstep:} Can string theory provide us us with the correct Compton amplitude for a classical Kerr black hole? Recently, it has been shown that all known Kerr scattering amplitudes are compatible with established tree-level unitarity constraints from the higher-spin literature~\cite{Chiodaroli:2021eug}. Higher-spin theories are close cousins of string theory and share many of the stringy features, which suggest synergies. The massive string modes are prime examples of how higher-spin states can be embedded in a unitary and mathematically consistent framework, and may be of use for modelling classical Kerr amplitudes. 

\smallskip

Since string theory provides a microscopic description of extremal black holes~\cite{Strominger:1996sh} it should provide a natural framework for exploring scattering amplitudes of such objects, also in the macroscopic limit. Observations of near-extremal Kerr black holes are expected from astrophysics, and it would be interesting to describe their scattering amplitudes, and more generally their waveforms, using results such as the Kerr/CFT correspondence~\cite{Guica:2008mu}.

\smallskip

While quantum and stringy effects for astrophysical black holes are expected to be very small, it nevertheless is an interesting question in principle how to incorporate such effects in scattering amplitudes and waveform calculations. In particular, properly describing Hawking radiation~\cite{Hawking:1975vcx} from an EFT~\cite{Goldberger:2020geb} or amplitudes perspective could lead to significant progress in the understanding of quantum gravity.

\section{Mathematical structures}

The interplay between duality, supersymmetry, and the low-energy expansion of superstring amplitudes has resulted in  the discovery of unexpected and exciting algebraic and arithmetic structures. Numerous mathematical properties of
string amplitudes are now understood in terms of recent developments in number theory, modular forms,  
and algebraic geometry. Conversely, the low-energy expansion of 
string amplitudes and their integrands has spawned new directions of mathematical research related to  single-valued integration and generalizations of non-holomorphic modular forms.

\smallskip

Already at genus zero, the pattern of multiple zeta values (MZVs) in higher-point functions is elegantly understood in terms of Hopf-algebra structures of motivic MZVs and the Drinfeld associator \cite{Schlotterer:2012ny, Drummond:2013vz}. As a result, tree-level amplitudes in general string theories furnish clean showcases of the coaction conjecture \cite{Schnetz:2013hqa, Francislecture} on the stability of amplitudes under the motivic coaction. So far, this coaction conjecture has found manifestations in $\phi^4$-theory \cite{Panzer:2016snt}, the magnetic moment of the electron \cite{Schnetz:2017bko}, ${\cal N}=4$ super-Yang--Mills \cite{Caron-Huot:2019bsq},  and various 
families of Feynman integrals \cite{Abreu:2017mtm, Brown:2019jng, Abreu:2021vhb}. Therefore, understanding the structure of string amplitudes may contribute to unravelling number-theoretic symmetries shared by quantum field theories and string theories.

\smallskip

Furthermore, the low-energy expansion of tree-level closed-string amplitudes  can be obtained from the low-energy expansion of open-string amplitudes through a formal operation on motivic MZVs \cite{Stieberger:2013wea, Stieberger:2014hba,Schlotterer:2018abc,Vanhove:2018elu,Brown:2019wna}, referred to as the single-valued map \cite{Schnetz:2013hqa, Brown:2013gia}. This process of inferring closed-string results from the open string goes considerably beyond the standard KLT formula of string theory \cite{Kawai:1985xq} and combinatorially realizes a field theory double copy \cite{Schlotterer:2012ny, Stieberger:2013wea} similar to that of open strings \cite{Zfunctions, Azevedo:2018dgo}. 
The single-valued map between tree-level amplitudes connects different perturbative string theories
beyond the scope of any known duality \cite{Stieberger:2014hba}, e.g.\ 
the gauge sectors of Heterotic and Type-I superstrings at weak coupling.

\smallskip

An ambitious long-term goal of both conceptual and technical appeal is to
reduce closed-string loop amplitudes to open-string building blocks.
The appearance of single-valued MZVs in closed-string tree-level amplitudes
found several echoes at genus one
\cite{Zerbini:2015rss, DHoker:2015wxz, Broedel:2018izr, Zagier:2019eus}
and led to a proposal for a single-valued map between configuration-space
integrals in open- and closed-string one-loop amplitudes \cite{Gerken:2020xfv}.
This proposal calls for extensions to the integration over modular parameters 
at genus one and to iterated integrals on higher-genus surfaces. The perturbative 
and possibly non-perturbative systematics of relating open- and closed-string interactions 
may have valuable lessons for old and new string dualities. 

\smallskip

Another mathematical viewpoint on open-closed-string relations that may advance 
loop-level developments is furnished by twisted de Rham theory. The $\alpha'$-dependent
Koba-Nielsen factor aligns moduli-space integrals of string tree amplitudes
into twisted cohomologies which have been studied in the mathematics literature of the 80's
and 90's. The KLT relations at tree-level then become a consequence of linear algebra
in these cohomologies \cite{Mizera:2017cqs} through the
so-called twisted period relations \cite{intersection}. The chiral variant of the KLT 
relations \cite{Huang:2016bdd} is another corollary of the twisted period relations
\cite{Mizera:2017rqa} that reproduces ambi-twistor-string amplitudes from a 
formal $\alpha' \rightarrow \infty$ limit \cite{Casali:2016atr}. 

\smallskip

Similarly, monodromy relations \cite{BjerrumBohr:2009rd, Stieberger:2009hq, 
Tourkine:2016bak, Hohenegger:2017kqy} among open-string amplitudes
descend from properties of twisted homologies \cite{Mizera:2019gea, Casali:2019ihm}. 
The latter also shed light on mixed open- and closed-string amplitudes in the gravity 
sector of Type I as exemplified in the recent one-loop discussion \cite{Stieberger:2021daa}, 
see \cite{Stieberger:2009hq, Stieberger:2015vya} for similar relations at tree-level.
Methods of twisted de Rham theory also became increasingly
important in particle physics, for instance for basis decompositions of Feynman
integrals \cite{Mastrolia:2018uzb, Abreu:2019wzk, Mizera:2019ose, Frellesvig:2020qot}.

\smallskip

At one loop and beyond, string amplitudes produce a wealth of elliptic and modular 
generalizations of MZVs. Closely related elliptic polylogarithms \cite{BrownLev} drive
the algorithmic integration of open-string insertions on a genus one worldsheet 
\cite{Broedel:2014vla, Broedel:2017jdo} and organize the
$\alpha'$-expansion of configuration-space integrals in terms of elliptic MZVs
\cite{Enriquez:Emzv}, see \cite{Mafra:2019xms, Broedel:2019gba} for recent 
all-order results. Closed-string 
applications of elliptic polylogarithms have been pioneered in \cite{Panzertalk, 
Broedel:2019tlz} and furnish rewarding targets for future research.
At the same time, elliptic polylogarithms and the associated iterated integrals
of modular forms substantially improved the computational 
reach for Feynman integrals beyond
genus-zero polylogarithms \cite{Adams:2017ejb, Broedel:2017kkb},
see in particular the references in the White Paper \cite{Bourjaily:2022bwx}.

\smallskip

Uniform transcendentality, familiar from dimensionally regularized Feynman 
integrals \cite{Kotikov:1990kg, Arkani-Hamed:2010pyv,Broedel:2018qkq}, enjoys a Type~I and Type~II 
superstring amplitude counterpart in their $\alpha'$-expansions at 
tree-level \cite{Henn:2013pwa, Broedel:2013aza} and at one loop 
\cite{Adams:2018yfj, Mafra:2019xms, DHoker:2019blr}.
However, uniform transcendentality of Type~II superstrings may get challenged 
at higher $\alpha'$-orders of one-loop amplitudes \cite{DHoker:2021ous} and may be 
violated in two-loop amplitudes \cite{Basu:2019fim}. Moreover, tree-level amplitudes of 
Heterotic and bosonic strings \cite{Huang:2016tag} violate uniform transcendentality.
This resonates with the comparison of loop amplitudes in different gauge and supergravity theories
whose transcendentality properties are sensitive to the amount of space-time supersymmetry \cite{Dunbar:1994bn, Boucher-Veronneau:2011rlc, Johansson:2014zca, Leoni:2015zxa,
Duhr:2019ywc, Kalin:2019vjc}.

\smallskip

The low-energy expansion of closed-string amplitudes at one-loop order and beyond has introduced fascinating families of non-holomorphic modular forms, dubbed modular graph functions \cite{Green:1999pv, Green:2008uj, DHoker:2015gmr, DHoker:2015wxz} and modular graph forms \cite{DHoker:2016mwo,DHoker:2016quv,DHoker:2019txf}, see \cite{Gerken:review} for an overview and \cite{Gerken:2020aju} for a {\sc Mathematica} package. Modular graph forms not only play a vital role in string-theory computations \cite{Basu:2016kli, Basu:2017nhs, Gerken:2018jrq, Basu:2019idd} but also
developed their own life in shaping mathematical research directions
\cite{DHoker:2017zhq,Brown:2017qwo,Brown:2017qwo2, Zagier:2019eus, Berg:2019jhh}. Their description 
via iterated integrals of holomorphic objects \cite{Broedel:2018izr, Gerken:2020yii}
exposes the intriguing web of differential and algebraic relations among modular graph 
forms discovered in \cite{DHoker:2015gmr,DHoker:2016mwo,DHoker:2016quv,Gerken:2018zcy}, sheds light on the interplay between open and closed strings \cite{Gerken:2020xfv} and
cross-fertilizes with studies of Poincar\'e series representations 
\cite{DHoker:2019txf, Dorigoni:2019yoq, Dorigoni:2021jfr}.

\smallskip

Similarly, building blocks of multi-loop closed-string amplitudes led to the
definition of higher-genus modular graph functions \cite{DHoker:2017pvk} 
with applications to the low-energy expansion of two-loop amplitudes
at four \cite{DHoker:2018mys} and five points \cite{DHoker:2020tcq}. In fact, 
the study of algebraic \cite{DHoker:2020tcq} and differential relations 
\cite{DHoker:2014oxd, Basu:2021xdt} among modular graph functions
beyond genus one motivates their generalization to modular graph 
tensors \cite{DHoker:2020uid}. Motivated in part by a construction of Kawazumi 
\cite{Kawazumi1, Kawazumi2}, modular graph tensors are introduced as non-holomorphic functions on 
the Torelli space of  compact Riemann surfaces of genus $g$ which transform as tensors under 
the modular group $Sp(2g,\mathbb Z)$.

\smallskip

{\bf \nextstep:} 
At higher genus, the study of modular graph functions \cite{DHoker:2017pvk} and tensors \cite{DHoker:2020uid} has just begun. A~natural first step is to explore  the differential structure of modular graph tensors in order to reveal the systematics of their algebraic relations, for instance via iterated-integral representations. This analysis should clarify the relations of modular graph tensors to non-holomorphic  Siegel modular forms  \cite{Pioline:2015qha} and improve the computational reach for multi-loop string amplitudes and their low-energy expansions.

\smallskip

The open-string analogues of modular graph tensors are uncharted terrain.
By analogy with genus one, the meromorphic building blocks of multiloop open-string
amplitudes are expected to inform the construction of higher-genus polylogarithms,
in particular the relevant classes of iterated integrals and their integration kernels. 
The meromorphic function spaces obtained from open-string input are likely to find 
applications to Feynman integrals in quantum field theory -- not only to compute 
further classes beyond elliptic polylogarithms but also as a case study in preparation 
for Feynman integrals associated with K3 or Calabi-Yau geometries and beyond.

\smallskip

The rich mathematical structures of string amplitudes and their moduli-space integrals are not fully manifest from their standard worldsheet prescriptions. Instead, the newly discovered mathematical structures might guide us to find radical reformulations of string perturbation theory from number-theoretic and algebraic-geometric principles, possibly without 
any reference to worldsheets. A grand goal would be to find a unified picture for the coaction properties of open-string tree-level  amplitudes \cite{Schlotterer:2012ny}, the derivation algebra governing modular graph  forms at genus one \cite{Gerken:2020yii} and similar structures to be found at higher genus.

\section{Concluding remarks}

In this White Paper, we have attempted to illustrate the richness of string perturbation theory through a lightning overview of recent progress and a small selection of future targets. Current developments in string amplitudes are not only strengthening the ties between different directions within string theory but are also triggering a striking confluence of ideas between string theorists, particle physicists, relativists, cosmologists, condensed matter physicists, and mathematicians. No doubt, the future targets listed in this White Paper merely reflect a tip of the iceberg of important technical and conceptual avenues for future research.

\section{Acknowledgements}

NB acknowledges partial financial support from CNPq grant number 311434/2020-7 and FAPESP grant numbers 2016/01343-7, 2019/24277-8 and 2019/21281-4. The research of ED is supported in part by the National Science Foundation under research grant PHY-19-14412. The research of MBG  has been partially supported by STFC consolidated grant ST/L000385/1.  The research of HJ is supported in part by the Knut and Alice Wallenberg Foundation under grants KAW 2018.0116 (From Scattering Amplitudes to Gravitational Waves) and KAW 2018.0162, and the Ragnar S\"oderberg Foundation (Swedish Foundations' Starting Grant). OS is supported by the European
Research Council under ERC-STG-804286 UNISCAMP.


%

\end{document}